# Enhanced Detective: A Dual-Stage Proactive Defense Against Internal Poisoning Attacks in Financial Fraud Detection Using Federated Learning

**Maria-Ioana-Andreea Ivan** [*] **Orhan Ermis** [**]
**Ruxandra F. Olimid** [*,***]

[*] *Department of Computer Science, University of Bucharest (e-mail: maria-ioana.ivan@s.unibuc.ro, ruxandra.olimid@fmi.unibuc.ro)*
[**] *Luxembourg Institute of Science and Technology (LIST), Luxembourg (e-mail: orhan.ermis@list.lu)*
[***] *Research Institute of the University of Bucharest (ICUB), Bucharest, Romania*

**Abstract:**
With recent developments in Artificial Intelligence (AI), the adoption of distributed learning paradigms such as Federated Learning (FL) has emerged. In simple terms, FL enables multiple entities to collaboratively train AI models without sharing local data. The FL paradigm is particularly valuable in privacy-sensitive settings, such as financial services, because it enables institutions to collaboratively improve fraud-detection models while keeping sensitive transaction data local. However, FL schemes are prone to security vulnerabilities, where malicious clients can poison local data or model updates and degrade the global model, and hence the performance of the resulting model might be affected. Therefore, in this work, we propose *Enhanced Detective*, a dual-stage robust aggregation strategy that combines Median Absolute Deviation (MAD), distance-based update filtering, and FedMedian aggregation. Experiments on an imbalanced credit card dataset evaluate targeted label-flipping and weight-scaling attacks over 50 communication rounds. With 40% malicious clients, *Enhanced Detective* achieves an AUPRC of 0.796 and a Recall of 0.847.

## 1. INTRODUCTION

As the digitalization of banking services continues to grow and the attack surface widens, the application of Machine Learning (ML) algorithms in the financial domain to detect suspicious activity is becoming a necessity. In contrast to traditional systems that rely on strict rules, modern models that use automated learning can adaptively extract complex correlations from large amounts of data, making it harder for attackers to bypass the system.

In the financial sector, dedicated laws and regulations such as the European Anti-Money Laundering Directives (AMLD) and the Payment Services Directive (PSD2) Commission (2024), as well as general requirements such as the GDPR Regulation (EU) 2016/679 (2016), prohibit the plain sharing of transactional data. In this context, Federated Learning (FL), a modern distributed learning approach that enables collaborative model training without the need for centralized training data, is particularly useful. FL reverses the traditional process, which brings the data to the model, and keeps sensitive data local by bringing the model to the data, sharing only the update of the model McMahan et al. (2017).

The present work examines how FL systems used for fraud detection respond to *internal poisoning* attacks, attacks for which internals corrupt the training data or the models themselves. Passive strategies mainly restrict the impact of malicious updates by simply isolating or dropping certain updates. However, they still miss some Byzantine nodes and do not allow them to participate in the training process. To overcome the limitations of passive robust aggregation methods, we propose a proactive defense strategy called *Enhanced Detective.* Our proposed method identifies malicious updates and then individually checks them, with the aim of dynamically reducing the impact of both extreme and more subtle malicious updates before performing the global anomaly detection process.

Our contribution can be summarized as follows:

- We propose *Enhanced Detective*, a novel proactive, dual-stage filtering mechanism for FL that mitigates internal poisoning attacks in financial fraud detection systems.
- We define our dual-stage *Enhanced Detective* strategy as an improved version of a simpler initial Detective strategy that proved inadequate under high compromise rates.

- We run comparative experiments on an imbalanced credit card fraud detection dataset across 50 communication rounds and show that *Enhanced Detective* achieves an AUPRC of 0.796 and a Recall of 0.847 under a cyber stress level of 40%.

The rest of the paper is organized as follows. Section 2 provides the background and places our work in relation to the existing literature. Section 3 gives the problem definition. Section 4 introduces our proposed solution. Section 5 presents the experimental results. Finally, the last section concludes.

## 2. BACKGROUND AND RELATED WORK

In this section, we first present the state of the art analysis with respect to the fraud detection using ML techniques considering effectiveness of different performance metrics used for interpreting the outputs of fraud detection models. Later, we will overview the robust aggregation approaches in the literature.

### *2.1 Fraud Detection*

The application of ML algorithms in the financial domain, especially for the detection of suspicious transactions, has become increasingly important as banking services continue to become more digital. In contrast to traditional systems, which are based on strict rules that can be easily bypassed, modern models based on automated learning have the ability to identify patterns in large datasets that are harder to bypass. For the classification of transactions, one of the most widely used feedforward neural network architectures is the Multi-Layer Perceptron (MLP) Goodfellow et al. (2016). The idea is therefore to use Rectified Linear Unit (ReLU) activation function since it helps the model converge faster during training, and the Sigmoid function is placed as the output layer of the neural network to produce a probability between 0 and 1, where 1 represents a fraudulent transaction, and 0 represents a legitimate one. The authors evaluated the resulting model using the Credit Card Fraud Detection dataset Machine Learning Group (2018). One of the challenges when working with fraud detection datasets is how greatly imbalanced they are. Usually, in these types of datasets, where the goal is to detect fraud, the fraudulent instances are represented by a very small proportion, around 0.2% or less Goodfellow et al. (2016). This high imbalance creates problems when training a model. Since most transactions are legitimate, the model can become biased towards the majority class. For instance, a model that classifies all transactions as legitimate could achieve an accuracy of 95%, while still missing almost all fraud cases. Such a model would therefore be useless from an operational point of view.

To overcome the challenge of imbalanced datasets, a central metric used is Recall, or sensitivity, defined as the ratio of true positive predictions to the total number of actual positive instances Davis and Goadrich (2006). The maximisation of Recall can lead to the minimisation of false negative cases, but relying only on Recall can result in an uncontrolled rise of false alarms, which would negatively affect user experience. Therefore, the Area Under the Precision-Recall Curve (AUPRC) is also used to evaluate the model's behaviour across all decision-making processes. In contrast to the Receiver Operating Characteristic (ROC) curve, which offers a more optimistic image of the imbalanced dataset, the Precision-Recall curve isolates the performance exclusively on the minority class Davis and Goadrich (2006). AUPRC measures the area under the curve while the classification threshold is changed from 0 to 1. A high AUPRC score demonstrates that the model manages to maintain high precision, meaning fewer false alarms, while also achieving high Recall and detecting most fraudulent transactions.

### *2.2 Robust Aggregation Schemes*

Although FL introduces privacy protection for the participating clients' local data locally to some extent, the aggregation process can be manipulated by malicious client behavior. For instance, poisoning attacks are used for manipulating either the local training data or the model updates transmitted by the clients to the aggregation server. Tolpegin et al. Tolpegin et al. (2020) presented data-poisoning attacks, where malicious clients modify their local class labels before training (data poisoning). Model-poisoning attacks, on the other hand, directly manipulate the client updates submitted to the aggregation server such as an attacker can scale its update to increase its influence on the resulting global model Karame et al. (2023). These studies demonstrate that the conventional averaging operation used by Federated Averaging (FedAvg) McMahan et al. (2017) can be vulnerable to manipulated model updates sent by clients. This motivates aggregation mechanisms that can limit the influence of anomalous client contributions.

To cope with the challenges of aggregation process, robust aggregation techniques replace FedAvg by employing additional statistical analysis, which are less sensitive to extreme values. For instance, FedMedian calculates the median of the received updates independently for each model coordinate, whereas Trimmed Mean removes extreme coordinate values before averaging the remaining values Yin et al. (2018). These strategies reduce the influence of outlying values but do not explicitly assess the consistency of each complete client update. Other robust aggregation strategies combine different forms of statistical analysis. Bulyan applies an existing Byzantine-resilient aggregation rule to construct a candidate set and then performs an additional coordinate-wise selection and averaging step El-Mhamdi et al. (2018). In contrast, Robust Federated Aggregation (RFA) uses the geometric median to aggregate complete client-update vectors, providing a multivariate alternative to coordinate-wise median aggregation Pillutla et al. (2022). FoolsGold addresses coordinated poisoning by comparing the similarity of clients' historical updates and reducing the influence of clients that exhibit highly similar behavior, as commonly observed in Sybil attacks Fung et al. (2020). These methods illustrate that robustness can be introduced through multi-stage aggregation, multivariate statistics, or client-behavior analysis. However, coordinated Sybil attacks are outside the threat model considered in this work.

Beyond individual aggregation algorithms, Ermis and Khadraoui Ermis and Khadraoui (2026) examined the broader problem of model-update trustworthiness in FL and proposed MultiTrust-FL, a multi-layered framework combining update anomaly detection, client reputation assessment, and defense mechanisms. That work provides a general framework for trustworthiness assurance, whereas the present study focuses on a concrete statistical filtering and aggregation strategy and evaluates it against targeted label-flipping and weight-scaling attacks in financial fraud detection. Additionally, update clipping mechanism can be used as an alternative approach to remove the influence of unusually large updates. It has also been employed together with filtering and robust aggregation mechanisms in FL Nguyen et al. (2022). Moreover, the median absolute deviation (MAD) provides a robust measure of statistical dispersion Leys et al. (2013), while Euclidean distance can be used to quantify the deviation of an update from a reference model Blanchard et al. (2017); Goodfellow et al. (2016).

Existing approaches either limit the influence of extreme values at the coordinate level or assess the deviation of complete client updates. However, distance-based filtering performed directly on unbounded updates may be affected by some malicious client updates, e.g. if they are scaled higher or lower than the average. This limitation motivates *Enhanced Detective*, which combines robust coordinate processing, update-level screening, and robust aggregation within a unified strategy.

## 3. PROBLEM DEFINITION

In this section, we formalize the problem studied with respect to the FL setting and the aggregation strategies used as baselines, specification of the threat model including the label-flipping and weight-scaling attacks considered and finally, the robust aggregation objective as limiting the influence of malicious client updates while preserving the predictive performance of the global fraud-detection model.

### *3.1 Federated Learning and Aggregation Strategies*

Federated Learning (FL) is a distributed learning paradigm that enables the training of a particular ML model with the involvement of distributed nodes McMahan et al. (2017). The generic FL architecture is as illustrated in Figure 1. The training process is coordinated by an aggregation server, which initializes a global model and distributes its parameters to the participating clients. Each client trains a local copy of the model using its private dataset and returns the resulting model parameters or updates to the server. The server combines the received contributions using an aggregation function and uses the result to update the global model. In Federated Averaging (FedAvg), the client models are aggregated through a weighted average, where each client's contribution is typically weighted according to the size of its local training dataset. This process is repeated over multiple communication rounds until a predefined convergence criterion or maximum number of rounds is reached.

More formally, let $S$ denote the aggregation server, $\mathcal{C} = \{C_1, C_2, \ldots, C_K\}$ the set of $K$ participating clients, and $w^t \in R^d$ the global model at communication round $t$, where $d$ denotes the number of model parameters. Then, the server distributes the current global model $w^t$ to the participating clients. Each client $C_i$ trains the received model using its local dataset and obtains a locally trained model $w_i^t$. The resulting model update is defined as $\Delta w_i^t = w_i^t - w^t$. The set of model updates received from the participating clients in round $t$ is denoted by

$$\Delta W^t = \left\{\Delta w_1^t, \Delta w_2^t, \ldots, \Delta w_K^t\right\}.$$

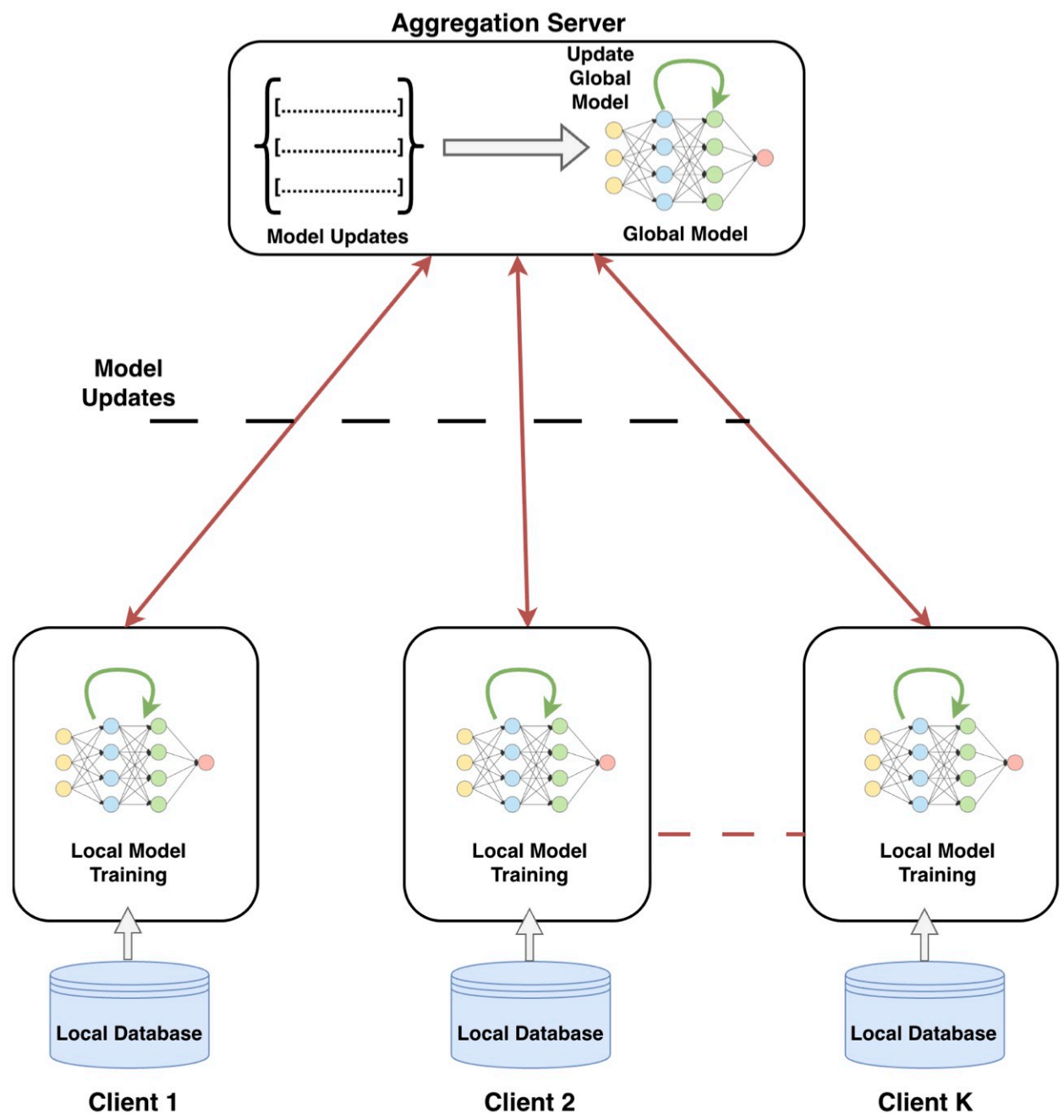

Fig. 1. Overview of a Federated Learning scheme.

The aggregation server applies an aggregation function $\mathcal{A}$ to the received client updates to obtain the aggregated model updates as follows:

$$\Delta w^t = \mathcal{A}\left(\Delta W^t\right).$$

The global model is subsequently updated as

$$w^{t+1} = w^t + \Delta w^t.$$

Therefore, $\mathcal{A}\left(\Delta W^t\right)$ represents an aggregated model update rather than a complete global model.

In Federated Averaging (FedAvg), the client updates are averaged according to the sizes of the clients' local datasets McMahan et al. (2017). Let $n_i$ denote the number of training samples held by client $C_i$, and let $n = \sum_{i=1}^{K} n_i$. The FedAvg aggregation function is defined as given below:

$$\mathcal{A}_{\text{FedAvg}}\left(\Delta W^t\right) = \sum_{i=1}^{K} \frac{n_i}{n} \Delta w_i^t.$$

Consequently, clients with larger local datasets have a greater contribution to the aggregated update.

FedMedian instead computes the median of the client updates independently for each model coordinate Yin et al. (2018). For each coordinate $j \in \{1, \ldots, d\}$, it is defined as follows:

$$[\mathcal{A}_{\text{FedMedian}}(\Delta W^t)]^j = \text{median}\left(\Delta w_1^{t,j}, \Delta w_2^{t,j}, \ldots, \Delta w_K^{t,j}\right),$$

where $\Delta w_i^{t,j}$ denotes coordinate $j$ of the update submitted by client $C_i$. By selecting the median coordinate value, FedMedian limits the influence of extreme individual values on the aggregated update.

Trimmed Mean is another coordinate-wise aggregation strategy Yin et al. (2018). For each coordinate $j$, let

$$\Delta w_{(1)}^{t,j} \leq \Delta w_{(2)}^{t,j} \leq \cdots \leq \Delta w_{(K)}^{t,j} \tag{1}$$

denote the ordered coordinate values received from the $K$ clients. Given a trimming parameter $b$ satisfying $0 \leq b < K/2$, Trimmed Mean removes the $b$ smallest and $b$ largest values and averages the remaining values:

$$[\mathcal{A}_{\text{TM}}(\Delta W^t)]^j = \frac{1}{K-2b}\sum_{r=b+1}^{K-b} \Delta w_{(r)}^{t,j}. \tag{2}$$

All the aggregation strategies, namely FedAvg, FedMedian, and Trimmed Mean defined in this section are therefore different in how they combine the received client updates. FedAvg computes a dataset-size-weighted average, whereas FedMedian and Trimmed Mean use coordinate-wise statistical operations that reduce the influence of extreme coordinate values. These aggregation functions provide the baseline aggregation strategies considered in this work, while FedMedian is also employed as the final aggregation operation of the proposed *Enhanced Detective* strategy.

### *3.2 Threat Model*

In this section, we introduce the threat model for the FL schemes. First, we assume that the aggregation server, $S$, is honest and fully follows the aggregation protocol, while a subset of the participating clients may behave maliciously. Therefore, for the set of participating clients, $\mathcal{C} = \{C_1, C_2, \ldots, C_K\}$, in the FL scheme, we assume that $\mathcal{C}_H$ and $\mathcal{C}_M$ correspond to the set of honest and malicious participants, respectively. We also assume that $\mathcal{C} = \mathcal{C}_H \bigcup \mathcal{C}_M$ and $\mathcal{C}_H \cap \mathcal{C}_M = \emptyset$ conditions hold for our setting. Honest clients follow the prescribed local training procedure and submit their resulting updates to the aggregation server. Malicious clients may manipulate their local training data or the model updates submitted to the server.

Let $\widehat{\Delta W}^t = \{\widehat{\Delta w}_1^t, \ldots, \widehat{\Delta w}_K^t\}$ be the snapshot of model updates transmitted to the aggregation server at communication round $t$, where $\widehat{\Delta w}_i^t$ denotes the update submitted by client $i$. For each honest client, $C_i$, the model update is defined as $\widehat{\Delta w}_i^t = \Delta w_i^t$, where $C_i \in \mathcal{C}_H$. On the other hand, we define the update submitted by malicious clients according to the label flipping attack and weight-scaling attack as follows.

*Label-flipping attack.* A malicious client manipulates its local training dataset before performing local training. In the considered fraud-detection task, the global model performs binary classification with class labels $\{0, 1\}$, where 0 represents a legitimate transaction and 1 represents a fraudulent transaction. In the label-flipping attack, a malicious client changes the labels of fraudulent transactions from 1 to 0 before performing local training. Thus, let $D_i$ denote the original local dataset of a malicious client $C_i \in \mathcal{C}_M$, and let $D_i^{\text{LF}}$ denote the corresponding poisoned dataset in which the labels associated with fraudulent transactions are changed from 1 to 0. The submitted model update is then calculated as

$$\widehat{\Delta w}_i^t = \text{LocalTrain}\left(w^t, D_i^{\text{LF}}\right) - w^t, \qquad C_i \in \mathcal{C}_M, \tag{3}$$

where $\text{LocalTrain}(w^t, D)$ denotes the local training procedure initialized with $w^t$ and performed using dataset $D$.

*Weight-scaling attack.* In this attack, a malicious client manipulates its locally computed model update before transmitting it to the aggregation server. Particularly, the client multiplies its update by a scaling factor, $\gamma > 1$, as given in the following equation:

$$\widehat{\Delta w}_i^t = \gamma \Delta w_i^t, \qquad \forall C_i \in \mathcal{C}_M. \tag{4}$$

The scaling operation preserves the direction of the original update while increasing its magnitude, thereby increasing its influence on the aggregated global update.

### *3.3 Robust Aggregation Objective*

Let $\widehat{\Delta W}^t = \{\widehat{\Delta w}_1^t, \ldots, \widehat{\Delta w}_K^t\}$ denote the set of client updates received by the aggregation server at round $t$. The objective is to construct a subset $\mathcal{C}_T^t \subseteq \mathcal{C}$ containing the clients whose updates are treated as trusted in that round. Based on the selected updates, the robust aggregated update is computed as follows:

$$\widehat{\Delta w}^t = \mathcal{A}_{\text{rob}}\left(\left\{\widehat{\Delta w}_i^t \mid C_i \in \mathcal{C}_T^t\right\}\right), \tag{5}$$

where $\mathcal{A}_{\text{rob}}$ denotes the robust aggregation function. The global model is subsequently updated as

$$w^{t+1} = w^t + \widehat{\Delta w}^t. \tag{6}$$

In the ideal setup, the robust aggregated update should remain as close as possible to the update that would be obtained using only the honest clients. Let this honest-only reference update be defined as

$$\Delta w_{\mathcal{C}_H}^t = \mathcal{A}_{\text{rob}}\left(\left\{\Delta w_i^t \mid C_i \in \mathcal{C}_H\right\}\right). \tag{7}$$

The aggregation error caused by malicious participation can then be expressed as

$$\mathcal{E}^t = \left\|\widehat{\Delta w}^t - \Delta w_{\mathcal{C}_H}^t\right\|_2, \tag{8}$$

where $\|\cdot\|_2$ denotes the Euclidean norm.

Because the identities of the malicious clients and the honest-only reference update are not directly available to the aggregation server, $\mathcal{E}^t$ cannot be evaluated during the aggregation process. The practical objective is therefore to limit the influence of suspicious updates while preserving the predictive performance of the global model. In the considered fraud-detection task, this performance is evaluated using the Area Under the Precision–Recall Curve (AUPRC) and Recall.

The central challenge is to construct a reliable trusted subset under different forms of malicious behavior. Large-magnitude updates produced by weight-scaling attacks can distort update-level distance measurements, whereas updates resulting from label-flipping attacks may exhibit less explicit deviations. A robust aggregation strategy should therefore limit the influence of both attack types without requiring prior knowledge of the malicious clients' identities.

## 4. PROPOSED SOLUTION

The proposed method *Enhanced Detective* not only identifies the malicious updates but also checks each one using a mathematical verification consisting of two consecutive filtering stages, as detailed in Algorithm 1. The main objective is to dynamically reduce the impact of both extreme and more subtle malicious updates before performing global anomaly detection.

**Phase 1.** In the first stage, all the updates received from all participating clients by the central server are processed coordinate by coordinate. For each parameter coordinate $w_j$ of the model, the algorithm computes the median value $med^j$ and the Median Absolute Deviation (MAD), denoted as $mad^j$ Leys et al. (2013):

$$mad^j = \text{Median}(\{|w_1^j - med^j|, ..., |w_K^j - med^j|\}) \quad (9)$$

Based on these statistical values, the algorithm creates dynamic limits that are updated during each round of training Nguyen et al. (2022). A small constant is also added for numerical stability, namely $\epsilon = 10^{-5}$, and the limits are defined as:

$$lower_bound = med^j - 5 \cdot mad^j - \epsilon \quad (10)$$

$$upper_bound = med^j + 5 \cdot mad^j + \epsilon \quad (11)$$

If a parameter goes outside the allowed range [$lower_bound$, $upper_bound$], it is moved back to the closest limit Nguyen et al. (2022). This reduces the impact of extreme values produced by model-poisoning attacks Karame et al. (2023) and prevents these abnormal updates from affecting subsequent detection steps. Finally, the parameters are flattened into one single vector for each client, denoted as $\tilde{w}_i$.

**Phase 2.** The server then calculates a global median vector, denoted as $v_{med}$, which serves as a reference because it is less affected by extreme values. After that, the Euclidean distance between each client vector and the global median vector is calculated using the $L_2$ norm Blanchard et al. (2017); Goodfellow et al. (2016). This distance is denoted as $d_i$ and is calculated as:

$$d_i = \|\tilde{w}_i - v_{med}\|_2, \quad (12)$$

where $\tilde{w}_i$ represents the sanitized weight vector of client $i$, $v_{med}$ is the global median vector, and $d_i$ represents the distance between the client update and the global median.

This second stage is useful for detecting label-flipping attacks, since these attacks can produce updates with magnitude similar to those of honest clients and may therefore pass the initial coordinate-wise clipping step Tolpegin et al. (2020).

Since suspicious updates are usually located farther from the main group of honest client updates, the algorithm applies MAD to the calculated distances as a second filtering step. The server then calculates a dynamic threshold using the median distance and its MAD value Leys et al. (2013):

$$threshold = \max(med_{dist} + 3 \cdot mad_{dist}, 1.5 \cdot med_{dist}), \quad (13)$$

where $med_{dist}$ represents the median of the calculated distances and $mad_{dist}$ represents their MAD. The threshold is set to the larger value between the MAD-based limit and 1.5 times the median distance.

**Phase 3.** If a client's distance exceeds this threshold, its update is considered suspicious and is excluded from the

---

**Algorithm 1** Enhanced Detective Strategy: Dual-Stage Filtering

**Require:** Client updates $W = \{w_1, w_2, ..., w_K\}$ from $K$ clients
**Ensure:** Aggregated global model $w_{global}$

**Phase 1: Coordinate-wise Clamping**
Let $\epsilon = 10^{-5}$ be a small constant for numerical stability
**for** each parameter coordinate $j$ **do**
  $med^j \leftarrow \text{Median}(\{w_1^j, w_2^j, ..., w_K^j\})$
  $mad^j \leftarrow \text{Median}(\{|w_1^j - med^j|, ..., |w_K^j - med^j|\})$
  $upper_bound \leftarrow med^j + 5 \cdot mad^j + \epsilon$
  $lower_bound \leftarrow med^j - 5 \cdot mad^j - \epsilon$
  **for** each client $i \in \{1, ..., K\}$ **do**
    $\tilde{w}_i^j \leftarrow \text{Clip}(w_i^j, lower_bound, upper_bound)$
  **end for**
**end for**
Flatten all sanitized updates into vectors $\tilde{W} = \{\tilde{w}_1, ..., \tilde{w}_K\}$
**Phase 2: Global Distance Filtering**
$v_{med} \leftarrow \text{Median}(\tilde{W})$
**for** each client $i \in \{1, ..., K\}$ **do**
  $d_i \leftarrow ||\tilde{w}_i - v_{med}||_2$
**end for**
$med_{dist} \leftarrow \text{Median}(\{d_1, ..., d_K\})$
$mad_{dist} \leftarrow \text{Median}(\{|d_1 - med_{dist}|, ..., |d_K - med_{dist}|\})$
$threshold \leftarrow \max(med_{dist} + 3 \cdot mad_{dist}, 1.5 \cdot med_{dist})$
**Phase 3: Robust Aggregation**
$Trusted \leftarrow \emptyset$
**for** each client $i \in \{1, ..., K\}$ **do**
  **if** $d_i \leq threshold$ **then**
    $Trusted \leftarrow Trusted \cup \{w_i\}$
  **end if**
**end for**
**if** $Trusted$ is empty **then**
  $Trusted \leftarrow W$
**end if**
$w_{global} \leftarrow \text{FedMedian}(Trusted)$
**return** $w_{global}$

**Algorithm 2** Initial Detective Strategy

**Require:** Client updates $W = \{w_1, w_2, \ldots, w_K\}$ from $K$ clients
**Ensure:** Aggregated global model $w_{global}$
**Phase 1: Vector Flattening and Reference Calculation**
**for** each client $i \in \{1, \ldots, K\}$ **do**
  $\tilde{w}_i \leftarrow \text{Flatten}(w_i)$
**end for**
$v_{med} \leftarrow \text{Median}(\{\tilde{w}_1, \ldots, \tilde{w}_K\})$
**Phase 2: Distance Calculation and Thresholding**
**for** each client $i \in \{1, \ldots, K\}$ **do**
  $d_i \leftarrow ||\tilde{w}_i - v_{med}||_2$
**end for**
$med_{dist} \leftarrow \text{Median}(\{d_1, \ldots, d_K\})$
$mad_{dist} \leftarrow \text{Median}(\{|d_1 - med_{dist}|, \ldots, |d_K - med_{dist}|\})$
$threshold \leftarrow \max(med_{dist} + 3 \times mad_{dist},\ 1.5 \times med_{dist})$
**Phase 3: Robust Aggregation - Same as in Algorithm 1**

current communication round. The remaining trusted updates are then aggregated using the FedMedian strategy, helping to keep the global model stable when malicious updates are present Yin et al. (2018).

To overcome the basic limitations of passive defenses, a simpler proposed proactive strategy, called *Detective*, was initially considered. This initial strategy relies on client isolation based on the Median Absolute Deviation Leys et al. (2013) and its two phases, which differ from those of the *Enhanced Detective*, are given in the algorithm. As discussed in the next section, Detective was a first version that led to the definition of the improved two-filtering-stage algorithm, *Enhanced Detective*.

## 5. EXPERIMENTAL RESULTS

This section compares the performance of the federated network under several attack scenarios. The primary objective is to demonstrate the drawbacks of conventional aggregation algorithms and to assess the performance of the proposed protection mechanisms. The research is based on measures related to the financial sector, such as AUPRC, Recall, and the Confusion Matrix Davis and Goadrich (2006). The findings are observed across 50 communication cycles to examine how the model's performance varies over time. The simulation environment was set up locally on a laptop with an AMD Radeon RX 7600S graphics card, 16 GB RAM and AMD Ryzen 9 7940HS CPU.

### *5.1 Baseline Performance and Vulnerability Analysis*

To establish a baseline reference point, initial tests evaluated the classic Federated Averaging algorithm in a completely honest environment McMahan et al. (2017). After 50 communication rounds, the classical FedAvg algorithm achieves an AUPRC score of about 0.86 and a Recall score of about 0.85. These findings were close to those produced by the centralized model and, in some cases, even significantly better. This indicates that in the absence of malicious clients on the network, federated learning can achieve similar performance while keeping clients' data local.

However, after the introduction of malicious clients, severe vulnerabilities were discovered. Under a data-poisoning attack, such as label flipping Tolpegin et al. (2020), where malicious nodes intentionally flip labels from 1 to 0, the Recall score of the FedAvg algorithm collapsed to roughly 0.23, failing to flag the majority of fraudulent transactions. Moreover, under a model-poisoning attack, namely Weight Scaling Karame et al. (2023), where attackers multiply their weights by an extreme factor, thereby attempting a DoS attack Nguyen et al. (2022), the system collapsed around round 10.

### *5.2 Evaluation of Passive Defenses*

To counteract the vulnerabilities found in the baseline algorithm, the conducted experiments evaluated the performance of two classical robust aggregation strategies, namely Federated Median and Trimmed Mean Yin et al. (2018); Blanchard et al. (2017). In the case of model-poisoning attacks, the FedMedian method showed very high resistance to severe weight scaling. This means that malicious clients sending extremely large values have little impact on the final aggregated model, as these values are viewed as outliers rather than updates that reflect the model Yin et al. (2018). Even at a compromise rate of 40%, FedMedian maintained the network stable, achieving a Recall score of approximately 0.78 and preventing the collapse of the network.

In contrast, Trimmed Mean highlighted the problem of dependence in estimating the attack rate Yin et al. (2018). Configured to eliminate extreme values from the distribution, this strategy functioned without problems under the 20% attack scenario. However, when the attack intensity increased to 40%, the algorithm mathematically failed. Since only a part of the malicious clients were removed, the remaining extreme vectors corrupted the final average, causing the system to collapse during the ninth communication round.

The label-flipping attack exposed a common weakness of the passive methods Tolpegin et al. (2020). Since compromised clients transmit weight changes that do not exhibit extreme anomalies but only direct the model towards incorrect decisions, the coordinate-wise filtering process becomes inefficient. Under a 40% attack scenario, Trimmed Mean showed a visible performance degradation, with the Recall score decreasing to approximately 0.50, the network failing to identify a significant number of fraudulent transactions. Although FedMedian performed better, maintaining a Recall score close to 0.70, it still registered a decrease in performance when compared to the ideal environment.

The major conclusion of this evaluation stage is that passive defenses strictly treat the symptom of the problem by restricting the impact of malicious updates, while still allowing Byzantine nodes to participate in the training process.

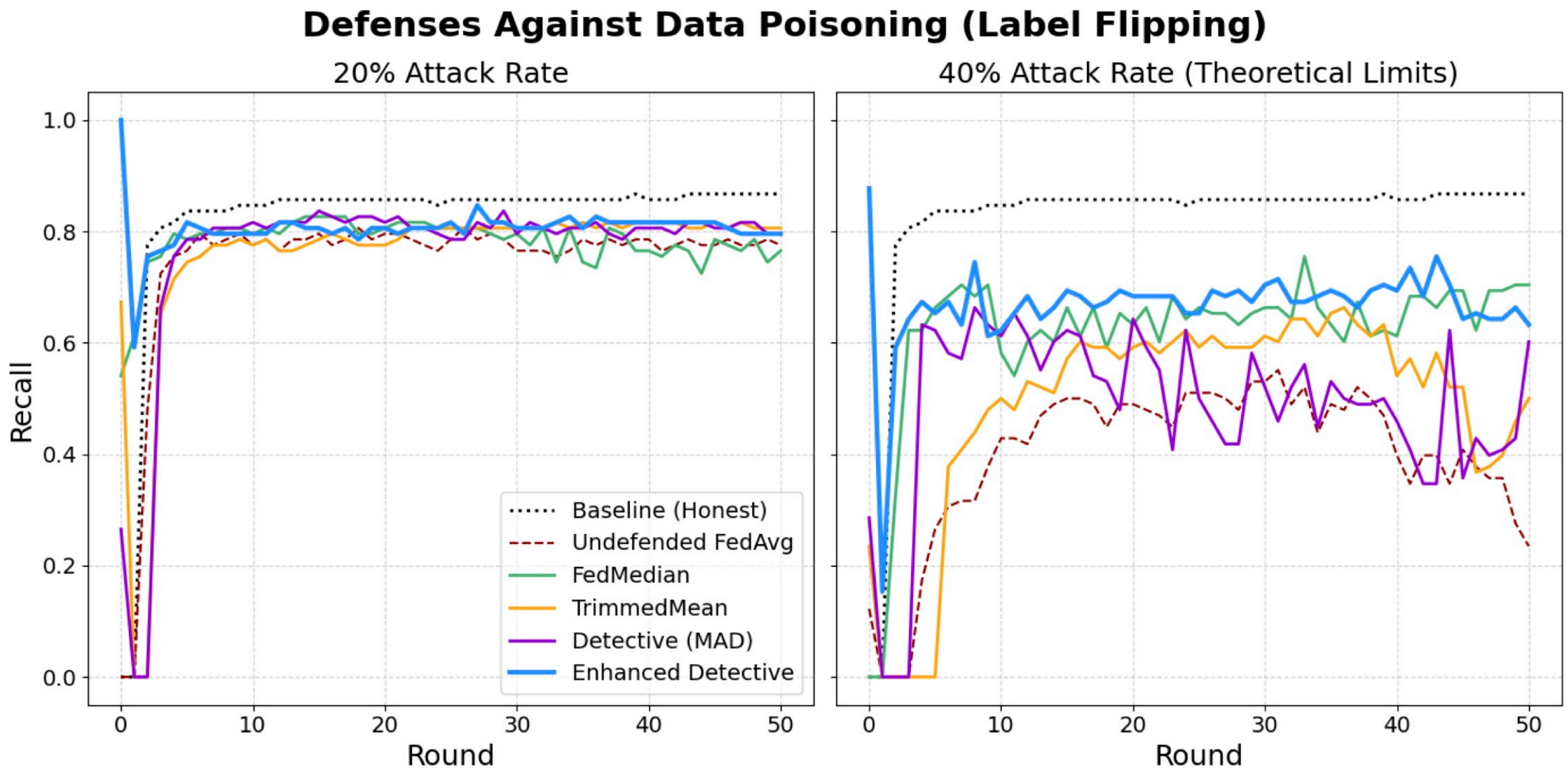

Fig. 2. The defense methods' performance against data poisoning.

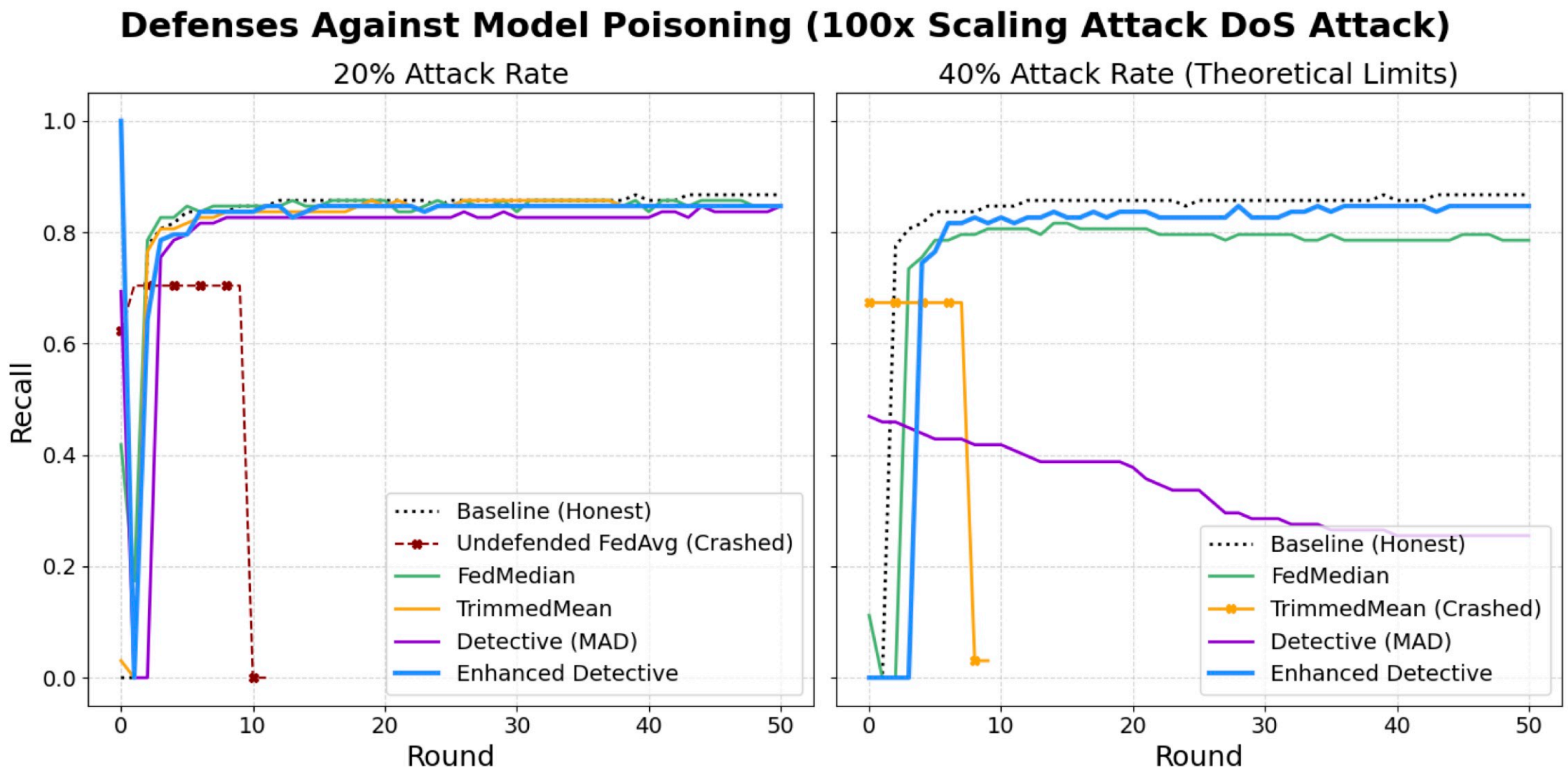

Fig. 3. The defense methods' performance against model poisoning.

### *5.3 Performance of the Proposed Proactive Defense*

To overcome the basic limitations of passive defenses, the experiments first evaluated the performance of an initially proposed proactive strategy, called Detective, based on client isolation using the Median Absolute Deviation Leys et al. (2013) and defined in the previous section.

The comparison graphs in Figure 2 and Figure 3 show that the first version of the Detective strategy performed very well against label-flipping attacks Tolpegin et al. (2020). It worked better than the Trimmed Mean method when the attack rate was set to 20%. This means it was able to identify and isolate the malicious nodes that were sending false updates and leading the model to make incorrect predictions. However, the ultimate stress test, namely model-poisoning Karame et al. (2023) at a compromise rate of 40%, achieved through weight scaling by a factor of 100x, revealed a structural vulnerability. Although the algorithm managed to prevent the complete collapse of the network, surviving until the 50th communication round in comparison with Trimmed Mean, which failed during round 9, the global $L_2$ geometric distances became so strongly distorted by the magnitude of the malicious vectors that the MAD mechanism became excessively sensitive, resulting in a final Recall score of only 0.25.

The identification of this mathematical distortion led to the development of *Enhanced Detective*, which implements a dual filtering mechanism, as explained in the previous section and depicted in Algorithm 1. The effect of this algorithmic optimization proved to be substantial, as demonstrated by the obtained visualizations. In the critical model-poisoning scenario at a compromise rate of 40%, while the other strategies degraded the detection performance, *Enhanced Detective* instantly neutralized the extreme scaling factor. The network protected by the new algorithm had a performance curve that almost exactly matched that of the ideal baseline environment. According to the final aggregated metrics, illustrated in Figure 4, the upgraded approach achieved an AUPRC of 0.796 and a Recall of 0.847 under a cyber stress level of 40%, correctly classifying 83 fraudulent transactions while missing only 15, as detailed in the confusion matrix in Figure 5.

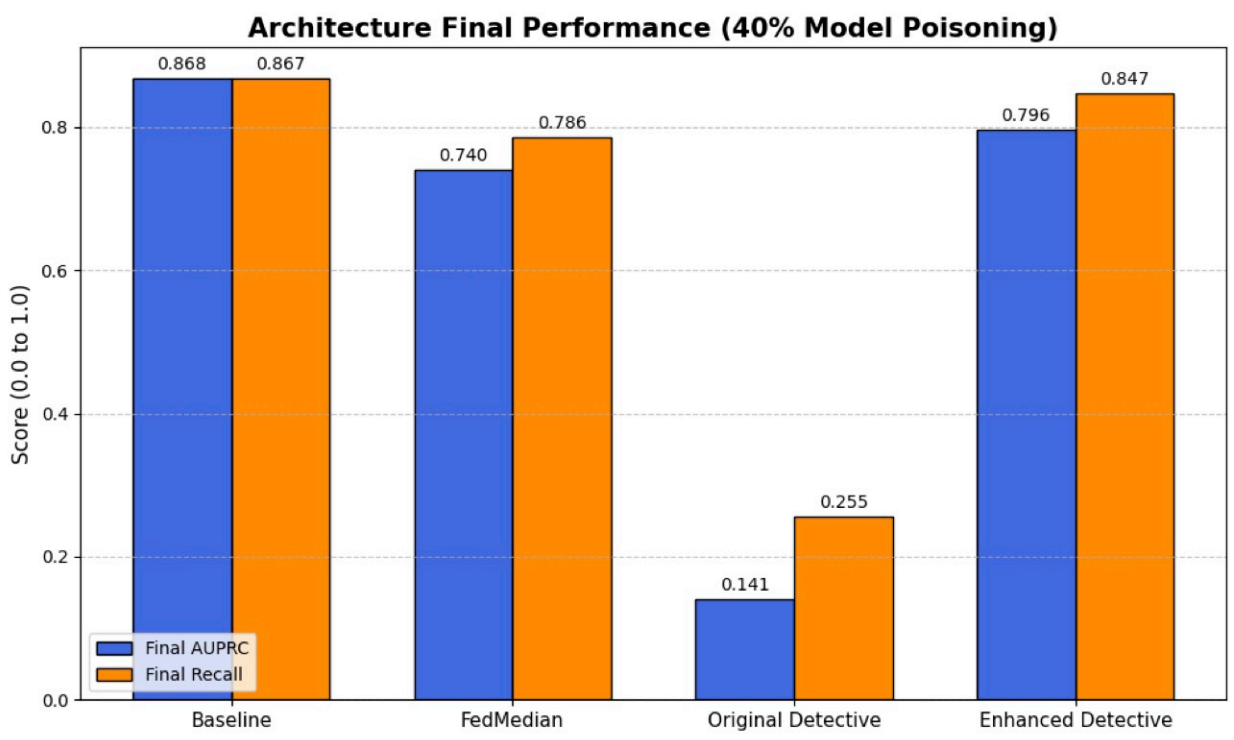

Fig. 4. The overall performance of the architecture.

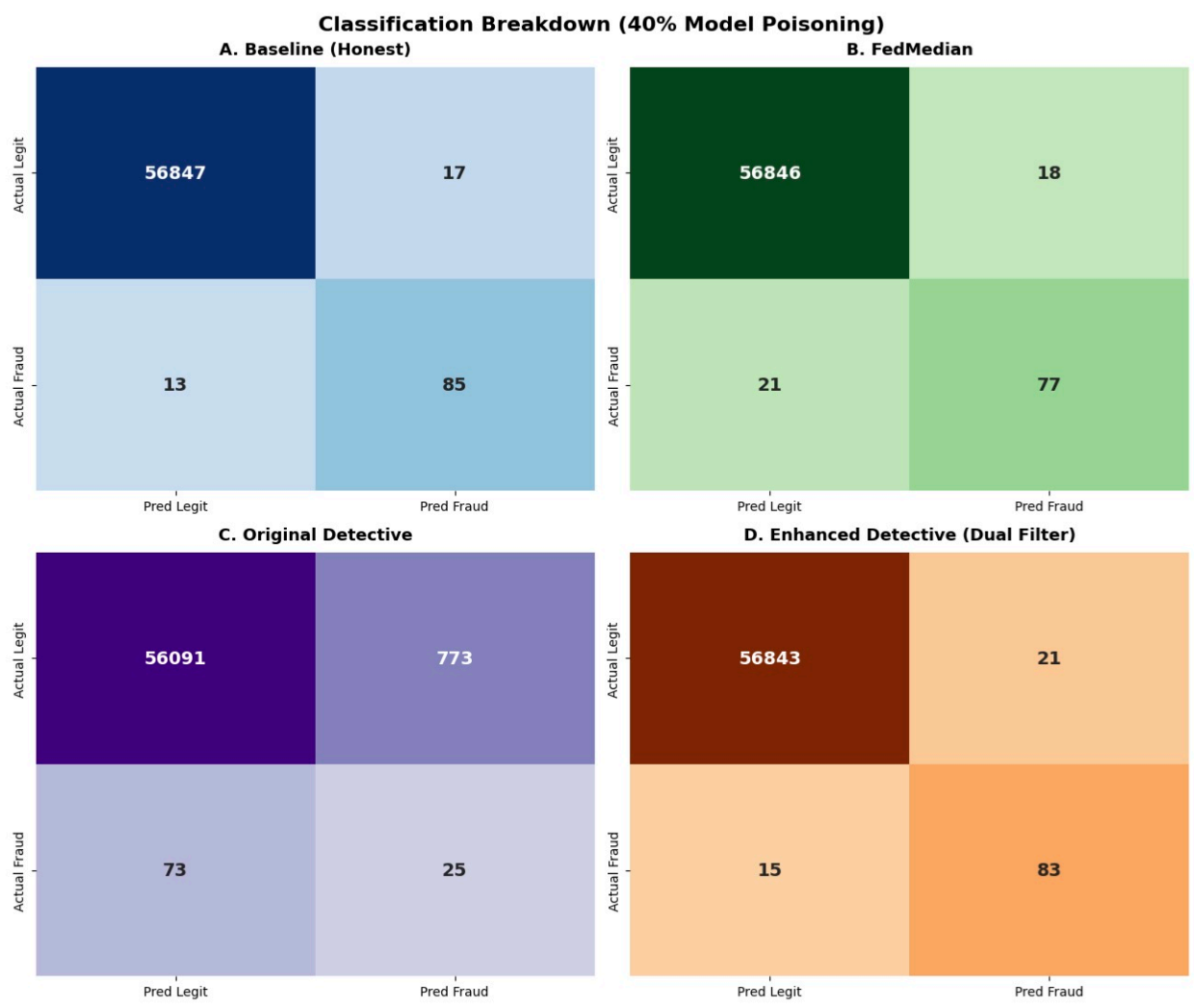

Fig. 5. Confusion Matrices comparing the final classification performance under a 40% model-poisoning attack.

## 6. CONCLUSION

We addressed FL's vulnerability to internal poisoning attacks in financial fraud detection and proposed *Enhanced Detective*, a proactive dual-stage filtering mechanism that isolates malicious updates prior to model aggregation. Our experimental evaluation on an imbalanced credit card fraud dataset across 50 communication rounds validated the efficacy of *Enhanced Detective* under extreme conditions (40% compromise rate), which is close to the ideal baseline environment, achieving an AUPRC of 0.796 and a Recall score of 0.847.

Future work might include a concrete evaluation of scalability with respect to the number of clients, as the proposed solution introduces additional time and resource costs that might limit real-time financial transaction systems. Further, we have considered identical settings; analyzing different transaction data among institutions remains a subject of future investigation. Finally, future work could extend the evaluation of *Enhanced Detective* to more evolved attacks, including adaptive behaviors and time-varying scaling strategies.

## ACKNOWLEDGMENTS

This work was supported by a grant of the Ministry of Research, Innovation and Digitalization, CNCS/CCCDI - UEFISCDI, project number ERANET-CHISTERA-IV-PATTERN, within PNCDI IV. This research was funded in part, by the Luxembourg National Research Fund (FNR), grant reference INTER/CHIST23/17931746. This work was mostly performed during M.I.A.Ivan MSc studies at the University of Bucharest.